\documentclass[10pt,twocolumn]{article}

\usepackage{graphicx} 
\graphicspath{{figures}} 
\usepackage[font=small,labelfont=bf]{caption} 

\usepackage{xcolor}
\definecolor{black}{HTML}{212427}
\definecolor{blue}{HTML}{0563C1}
\definecolor{red}{HTML}{B51700}

\usepackage{hyperref,nameref}
\hypersetup{colorlinks=true,allcolors=blue}
\newcommand{\rref}[2]{\hyperref[#1]{\ref{#1}#2}} 

\color{black}
\usepackage{setspace} 
\usepackage[margin=0.67in]{geometry} 
\usepackage[switch]{lineno} 
\usepackage{anyfontsize}
\usepackage{lipsum}
\usepackage{fancyhdr,lastpage}
\usepackage{titlesec}
\titleformat{\section}{\Large\bfseries}{\thesection. }{0mm}{}
\titleformat{\subsection}{\bfseries}{\thesubsection. }{0mm}{}
\titlespacing{\section}{0pt}{\baselineskip}{0pt}
\titlespacing*{\section}{0pt}{\baselineskip}{0pt}
\titlespacing*{\subsection}{0pt}{\baselineskip}{0pt}

\usepackage{amsmath,amssymb}
\renewcommand{\v}[1]{\boldsymbol{\mathbf{#1}}} 
\renewcommand{\t}[1]{\text{#1}} 

\usepackage{multirow} 

\usepackage{soul} 
\setstcolor{red}
\setul{}{0.7pt}

\usepackage[
  backend=biber,        
  autocite=superscript, 
  sorting=none,         
  style=numeric-comp,   
  maxnames=100,         
  minnames=100,         
  giveninits=false,     
  autopunct=false,      
  doi=true,             
  hyperref=true         
]{biblatex}
\DeclareFieldFormat{labelnumberwidth}{\;[#1]\!\!\!\!} 
\renewbibmacro{in:}{} 
\AtNextBibliography{\small} 
\AtEveryBibitem{\clearfield{pages}} 
\AtEveryBibitem{\clearfield{volume}} 
\AtEveryBibitem{\clearfield{number}} 
\AtEveryBibitem{\clearfield{month}} 

\usepackage{graphicx}
\usepackage{multirow}
\usepackage{booktabs}
\usepackage{array}

\begin{document}

\twocolumn[
  \begin{center}
    \large
     \textbf{Dislocation-mediated short-range order evolution during thermomechanical processing}
  \end{center}
  Mahmudul Islam$^1$,
  Killian Sheriff$^1$, and
  Rodrigo Freitas$^1${\footnotemark[1]} \\
  $^1$\textit{\small Department of Materials Science and Engineering, Massachusetts Institute of Technology, Cambridge, MA, USA} \\

  {\small Dated: \today}
  
  \vspace{-0.15cm}
  \begin{center}
    \textbf{Abstract}
  \end{center}
  \vspace{-0.35cm}
  Thermomechanical processing alters the microstructure of metallic alloys through coupled plastic deformation and thermal exposure, with dislocation motion driving plasticity and microstructural evolution. Our previous work\autocite{islam_nonequilibrium_2025} showed that the same dislocation motion both creates and destroys chemical short-range order (SRO), driving alloys into far-from-equilibrium SRO states. However, the connection between this dislocation-mediated SRO evolution and processing parameters remains largely unexplored. Here, we perform large-scale atomistic simulations of thermomechanical processing of equiatomic TiTaVW to determine how temperature and strain rate control SRO via competing creation ($\Gamma$) and annihilation ($\lambda$) rates. The simulations employ systems containing 2.4 million atoms and utilize a machine learning interatomic potential optimized to capture chemical complexity through the motif-based sampling technique. Using information-theoretic metrics, we quantify that the magnitude and chemical character of SRO vary systematically with processing parameters. We identify two regimes: a low-temperature regime with weak strain-rate sensitivity, and a high-temperature regime in which reduced dislocation density and increased screw character amplify chemical bias and accelerate SRO formation. The resulting steady-state SRO is far-from-equilibrium and cannot be produced by equilibrium thermal annealing. Together, these results provide a mechanistic and predictive link between processing parameters, dislocation physics, and SRO evolution in chemically complex alloys.
\vspace{0.2cm}


  \vspace{0.4cm}
]
{
  \footnotetext[1]{Corresponding author (\texttt{rodrigof@mit.edu}).}
}


\section{Introduction}

Thermomechanical processing (TMP) is a widely employed manufacturing technique in which alloys are strained at controlled temperatures to induce microstructural evolution via plastic deformation\autocite{rackwitz_understanding_2025, trink_processing_2023, yamanaka_stacking-fault_2017, sun_thermomechanical_2016, schuh_mechanical_2015,shahmir_effect_2016}. Plastic deformation is most often governed by dislocation motion; therefore, the mechanical response of a material undergoing TMP reflects dislocation behavior. That behavior is controlled by temperature and strain rate --- the two primary parameters of TMP\autocite{rizzardi_mild--wild_2022,kubilay_high_2021, lu_relative_2021}. For example, at low temperatures, dislocation motion is primarily governed by the Peierls barrier, whereas at high temperatures dislocations become thermally activated and can overcome this barrier more readily, reducing the flow stress at a fixed strain rate\autocite{marian_dynamic_2004}. Furthermore, increasing temperature activates qualitatively different modes of dislocation motion such as diffusion-assisted climb, and cross-slip --- which are largely suppressed at low temperatures\autocite{bulatov_computer_2006}. High strain rates, on the other hand, raise the flow stress so that dislocations surmount Peierls barriers with reduced thermal assistance (i.e., athermal glide). Increasing strain rate also promotes dislocation multiplication in order to sustain the high rate of plastic deformation\autocite{fan_strain_2021}. This leads to the formation of dense dislocation forests and junctions, many of which are sessile, that hinder the motion of glissile dislocations, resulting in strain hardening.

Fundamentally, dislocation motion is a bond-breaking and bond-forming process that alters the local chemical environment in the wake of moving dislocation cores. A large number of dislocations traversing through the system can thus modify the relative population of certain local chemical motifs, thereby altering spatial chemical correlations at the nanoscale --- commonly referred to as chemical short-range order (SRO) \autocite{sheriff_quantifying_2024, sheriff2024chemicalmotif, zhou_atomic-scale_2022}. SRO has been reported to influence several materials properties, including mechanical strength\autocite{antillon_chemical_2021,antillon_chemical_2020,dasari_exceptional_2023,chen_simultaneously_2021,zheng_multi-scale_2023}, radiation damage resistance\autocite{su_enhancing_2023,el_atwani_quinary_2023}, catalysis\autocite{xie_highly_2019}, and corrosion resistance\autocite{qiu_corrosion_2017,xie_percolation_2021, blades_tuning_2024}, especially in alloys with pronounced chemical complexity, such as high-entropy alloys\autocite{george_high-entropy_2019}. In ref.~\cite{islam_nonequilibrium_2025}, we demonstrated that dislocation motion both creates and destroys SRO, driving alloys into far-from-equilibrium SRO states. Notably, even alloys initialized as chemically random solid solutions evolve SRO under plastic deformation. The mechanism underlying this phenomenon was attributed to the chemically biased nature of dislocation motion in alloys, which promotes certain motifs over others\autocite{yin_atomistic_2021, utt_origin_2022, zhou_hidden_2021}. As described earlier, dislocation behavior is strongly dependent on temperature and strain rate. Consequently, a natural correlation must exist between these TMP parameters and the evolution of SRO via dislocation motion. However, this relationship remains obscure in the literature, leaving a critical gap in our understanding of how SRO evolves in metallic alloys. Without a mechanistic understanding of how dislocation motion affects SRO under different TMP conditions, interpretations of deformation experiments risk conflating intrinsic mechanical response with concurrent changes in SRO, and models cannot reliably predict regimes where nonequilibrium SRO dominates. This gap limits transferability across alloys and invites equilibrium-based extrapolations\autocite{moniri_three-dimensional_2023,chen_direct_2021, zhang_short-range_2020} that are systematically misleading for TMP.

Here, we investigate the effects of temperature and strain rate on the evolution of SRO in equiatomic TiTaVW alloy during TMP using large-scale atomistic simulations. These simulations --- performed on the Frontier exascale supercomputer\autocite{atchley_frontier_2023} --- employ a high-fidelity machine learning interatomic potential that was trained on ab initio data tailored to capture the structure and chemistry relevant to TMP (section~\ref{meth:1}). We select equiatomic TiTaVW as a canonical refractory alloy with a high Peierls barrier and screw-dominated plasticity, providing a stringent testbed for chemically-biased dislocation motion. In sections~\ref{sec:1}--\ref{sec:3} we demonstrate how TMP processing parameters influence SRO evolution. The dislocation motion mechanisms underlying SRO evolution are discussed in section~\ref{sec:4}. Finally, in section~\ref{sec:5}, we implement an empirical model based on our simulation results to explore steady-state SRO under conditions relevant to practical manufacturing processes.


\section{Methods}

\subsection{Machine learning interatomic potential}
\label{meth:1}

To conduct accurate atomistic simulations of equiatomic TiTaVW alloy, we developed a machine learning interatomic potential tailored to capture the chemistry-sensitive mechanisms governing dislocation behavior under thermomechanical loading --- mechanisms that conventional empirical potentials fail to represent accurately\autocite{sheriff_quantifying_2024,cao_capturing_2024}. The training dataset for the potential included both perfect and defect configurations, carefully curated to capture the chemical and structural complexity relevant to the TMP of TiTaVW alloy. The perfect configurations included body-centered cubic (bcc), face-centered cubic (fcc), and hexagonal close-packed (hcp) solid solutions, pure elemental bcc structures, and B2 structures representing all binary combinations. The fcc and hcp phases were included to account for phase stability, with fcc supercells specifically added due to their demonstrated ability to accurately capture screw dislocation core characteristics\autocite{Zotov_2024, wang2022tamingscrewdislocationcores}.

Defect configurations were chosen based on well-established mechanisms governing plastic deformation in bcc systems, which is primarily mediated by the motion of $\frac{1}{2} \langle 111 \rangle$ screw dislocations. The properties of these dislocations are strongly influenced by the stacking-fault energy (SFE) landscapes on the $(110)$ and $(112)$ planes\autocite{hirth1983theory, vitek1968intrinsic}. Accordingly, we included configurations with stacking faults along the $[111]$ direction on these planes. However, SFE curves alone are insufficient to fully represent the highly anisotropic core structures of dislocations in bcc systems\autocite{freitas_machine-learning_2022}. To address this, we incorporated configurations containing $\frac{1}{2} \langle 111 \rangle \{110\}$ screw dislocations generated via the dislocation dipole method\autocite{yin_ab_2020}, as well as configurations with $\frac{1}{2} \langle 111 \rangle \{110\}$ edge dislocation dipoles for their relevance in bcc plasticity\autocite{lee_strength_2021}. Finally, we included bcc configurations with both vacancies and interstitials (placed at either tetrahedral or octahedral sites) to account for defect-mediated mechanisms relevant to dislocation motion.

For all solid-solution phases, elemental compositions were sampled across the full TiTaVW local motif space using the motif-based algorithm of ref.~\cite{sheriff_machine_2025}. To bias the training dataset toward near equiatomic compositions, the procedure was adapted by first generating a diverse set of chemically random solutions and then allowing extra-cell swaps between configurations. This allowed greater flexibility to increase motif diversity, especially in low-data regimes, while favoring near equiatomic configurations to be represented.

The lattice constants of these structures were adjusted using Vegard's law to the sampled compositions. To mimic thermal vibrations, atomic displacements were applied by randomly perturbing atoms following the methodology in ref.~\cite{sheriff_machine_2025}. Isotropic lattice expansion was further applied to imitate both room-temperature and high-temperature thermal expansion\autocite{cao_capturing_2024}.

Density functional theory (DFT) calculations were performed to obtain energies and atomic forces for all configurations in the dataset (see Supplementary Information table 1). These calculations used the Perdew-Burke-Ernzerhof (PBE) exchange-correlation functional\autocite{PBE} and projector-augmented wave (PAW) pseudopotentials\autocite{PAW}, as implemented in VASP\autocite{VASP_1,VASP_2,VASP_3,VASP_4,VASP_PAW}. The resulting DFT energies and forces were used to train a machine learning interatomic potential based on the atomic cluster expansion framework, with hyperparameters optimized to balance both accuracy and generalization\autocite{sheriff_machine_2025,drautz_atomic_2019, lysogorskiy_performant_2021}.

All molecular dynamics and Monte Carlo simulations reported here were conducted using the developed machine learning interatomic potential. For more details on the training dataset structures and their construction, thermal noise and expansion augmentations, DFT simulation setup, and potential performance, we refer the reader to Supplementary Information section 1.

\subsection{Thermomechanical processing simulations}
\label{meth:2}

The central simulations underpinning the results of this paper were molecular dynamics simulations of uniaxial compressive deformation on bcc single crystals of TiTaVW alloy using the Large-scale Atomic/Molecular Massively Parallel Simulator (LAMMPS)\autocite{thompson_lammps_2022}. Simulation cells were initialized by randomly populating bcc lattice sites with Ti, Ta, V, and W atoms in equiatomic proportions. The initial simulation box measured $53a_\t{o} \times 106a_\t{o} \times 212a_\t{o}$ along the $[100]$, $[010]$, and $[001]$ directions, respectively, where $a_\t{o}$ is the temperature-dependent lattice constant. Each configuration contained approximately 2.4 million atoms, with periodic boundary conditions applied in all directions.

\begin{figure*}[!tb]
  \centering
  \includegraphics[width=\textwidth]{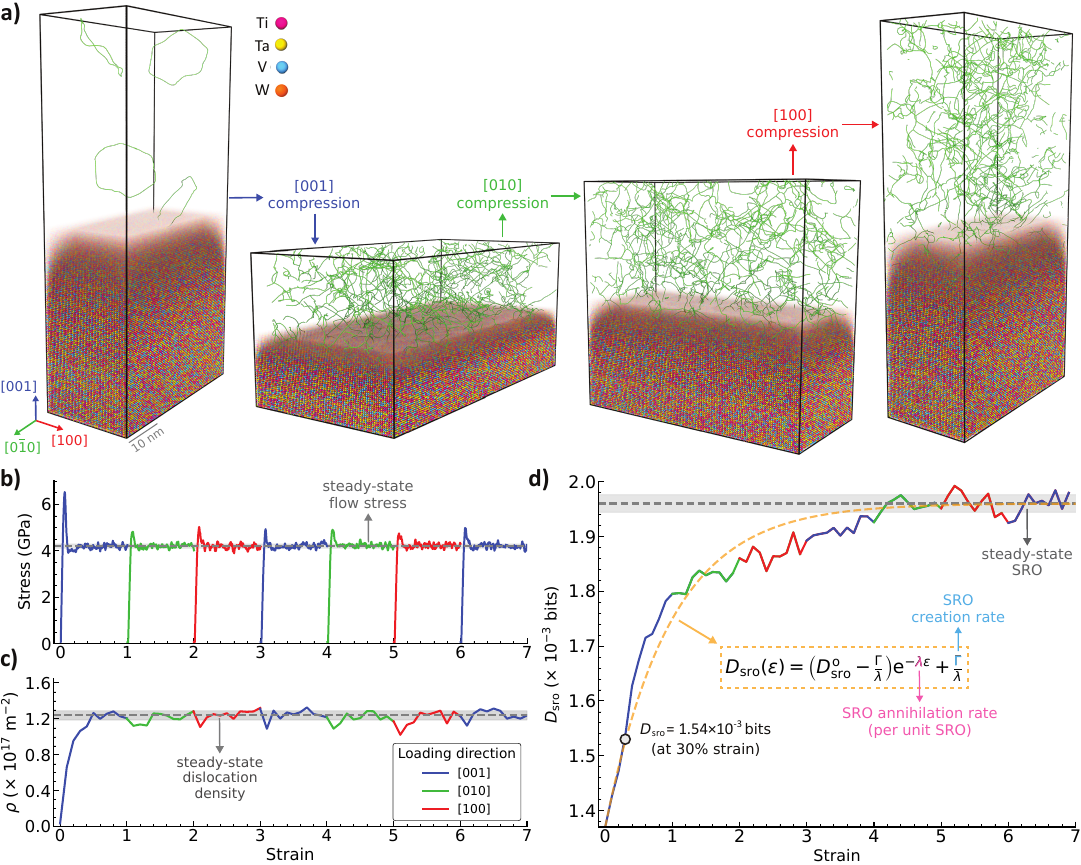}
  \caption{\label{figure_1} \textbf{Thermomechanical processing of equiatomic TiTaVW alloy.} \textbf{a)} Uniaxial compression of TiTaVW at $300\,\t{K}$ under a constant strain rate of $10^{9}\,\t{s}^{-1}$. After each unit strain, the deformation axis is rotated to align with the direction of largest dimension. Dislocation networks are shown as green lines. \textbf{b)} Stress-strain curve shows the onset of steady-state flow stress. \textbf{c)} Dislocation density also reaches steady-state with continued deformation. \textbf{d)} Evolution of SRO ($D_{\t{sro}}$) with strain, eventually reaching a steady-state SRO ($D^\infty_{\t{sro}}$). The dashed orange curve shows the fit to eq.~\ref{eq:sro}, capturing the observed $D_{\t{sro}}$ evolution. During fitting, $D^{\t{o}}_{\mathrm{sro}}$ was fixed at $D_{\mathrm{sro}}(\epsilon=0)$.}
\end{figure*}

To introduce initial plasticity carriers, four hexagonal, vacancy-type prismatic dislocation loops with a diameter of 10\,nm were embedded in the system. Each loop was assigned one of the four Burgers vectors of type $\frac{1}{2} \langle 111 \rangle$, resulting in an initial dislocation density of $3.6 \times 10^{15}\,\t{m}^{-2}$. The loops were placed randomly while avoiding overlaps, as illustrated in fig.~\rref{figure_1}{a}.

The dislocation-seeded structures were first equilibrated at constant temperatures $T$ ranging from 300\,K to 2,300\,K and zero pressure using a Nosé-Hoover thermostat and a Nosé-Hoover barostat, respectively, for 100\,ps, allowing isotropic volume relaxation. After equilibration, mechanical deformation\autocite{zepeda-ruiz_atomistic_2021,zepeda-ruiz_probing_2017} was applied in compression along the $[001]$ axis at constant strain rates $\dot{\epsilon}$ ranging from $10^8\,\t{s}^{-1}$ to $10^9\,\t{s}^{-1}$, while maintaining zero lateral pressure via the Nosé-Hoover barostat in the $[100]$ and $[010]$ directions. Temperature, which remained the same as the equilibration temperature, was controlled using a Nosé-Hoover thermostat throughout deformation. For all thermostats and barostats applied throughout the simulations, we used a damping parameter of 10\,ps and a chain length of 3.

After every unit strain, the compression axis was rotated sequentially from $[001]$ to $[0\bar{1}0]$, then to $[100]$, and again back to $[001]$. This procedure was repeated until a total true strain of 7 was reached. This cyclic deformation process (the first three cycles of which are illustrated in fig.~\rref{figure_1}{a}) mimics a ``metal kneading'' protocol\autocite{zepeda-ruiz_probing_2017}. The crystal, initially shaped with a dimension ratio of 1:2:4, is compressed along its longest axis, and the process is repeated along the new longest dimension in each subsequent cycle, preserving nearly constant volume and minimizing artifacts due to periodic boundary conditions. A full deformation sequence is provided in Supplementary Video 1. The integration timestep of the simulations was set to 5\,fs, which is $\sim$1/37th of the period of the highest-frequency phonon mode of equiatomic TiTaVW at 300K.

Dislocations in the systems were analyzed using the Dislocation Extraction Algorithm (DXA)\autocite{stukowski_extracting_2010}. Prior to dislocation analysis, the atomic structures were denoised using a score-based denoising algorithm\autocite{hsu_score-based_2024} for 8 steps. The full dislocation evolution during a simulation is shown in Supplementary Video 2. The cumulative edge-to-screw character ratio of dislocations in each simulation was quantified using the following expression:
\begin{equation}
    \label{eq:edge}
    R = \frac{\sum_i |\sin(\theta_i)| L_i}{\sum_i |\cos(\theta_i)| L_i},
\end{equation}
where $\theta_i$ is the angle between the line direction $\v{\ell}_i$ and the Burgers vector $\boldsymbol{b}_i$ for dislocation segment $i$, and $L_i$ is the length of that segment. In essence, $R$ represents a length-weighted ratio of edge to screw character across all dislocation segments in the system.

All visualizations were performed using the Open Visualization Tool (OVITO)\autocite{stukowski_visualization_2009}.

\subsection{Annealing simulations}
\label{meth:3}

To compare the SRO states obtained from TMP simulations with those resulting from equilibrium processing pathways (i.e., annealing), we performed Monte Carlo simulations to thermally equilibrate chemical configurations of equiatomic TiTaVW over a wide temperature range. Simulations were conducted from $500\,\t{K}$ to $10{,}000\,\t{K}$ in $500\,\t{K}$ increments. At each temperature, 65 independent simulations were performed, each containing $4{,}394$ atoms arranged in a cubic bcc cell of size $13a_\t{o} \times 13a_\t{o} \times 13a_\t{o}$ oriented along the $[100]$, $[010]$, and $[001]$ directions, where $a_\t{o}$ is the temperature-dependent lattice parameter. For temperatures above the melting point, $a_\t{o}$ was determined by extrapolating its linear dependence on temperature. Periodic boundary conditions were applied in all directions.

Simulations were initialized with a random equiatomic distribution of atoms on the bcc lattice. Thermal equilibration was achieved through atom-swap attempts between atoms of different elements, with acceptance governed by the Metropolis criterion\autocite{metropolis_equation_1953,hastings_monte_1970} probability $\exp(-\Delta E / k_\t{B}T)$, where $\Delta E$ is the energy change associated with the swap, $k_\t{B}$ is the Boltzmann constant, and $T$ is the temperature. Each simulation was carried out for 360,000 steps (i.e., 80 swaps per atom), with the first 180,000 steps used for initial equilibration. Seven statistically uncorrelated configurations were sampled every 30,000 steps during the remaining 180,000 steps, yielding equilibrium SRO states analogous to those obtained from thermal annealing.

\subsection{SRO quantification}
\label{meth:4}

To quantify chemical SRO in the states obtained from both TMP and annealing simulations, we employed the chemical motif framework introduced in refs.~\cite{sheriff_quantifying_2024} and \cite{sheriff2024chemicalmotif}. In this approach, the local chemical environment of an atom was described by a motif $\mathcal{M}_i$, consisting of the atom and its first-nearest neighbors. We utilized the Euclidean graph neural network architecture\autocite{smidt_euclidean_2021} from ref.~\cite{sheriff2024chemicalmotif} to identify all motifs $\mathcal{M}_i$ in a system and compute their population density distributions $P(\mathcal{M}_i)$, hereafter denoted simply as $P$ for brevity. The difference in SRO between two states, characterized by motif population density distributions $P_1$ and $P_2$, was quantified using the Jensen-Shannon divergence:
\begin{equation}
    \label{eq:js}
    D_{\t{JS}}(P_1 \parallel P_2) = \frac{1}{2} D_{\t{KL}}(P_1 \parallel M) + \frac{1}{2} D_{\t{KL}}(P_2 \parallel M),
\end{equation}
where $M = \frac{1}{2}(P_1 + P_2)$ and $D_{\t{KL}}$ denotes the Kullback-Leibler divergence\autocite{mackay2003information}. Larger values of $D_{\t{JS}}(P_1 \parallel P_2)$ indicate greater differences between $P_1$ and $P_2$.

While eq.~\ref{eq:js} can be used to compare two SRO states, it is often also desirable to quantify the absolute amount of SRO present in a single state. Here, this will be performed by measuring the divergence between the state of interest and a chemically random solid solution. For a state with motif population distribution $P$ this is computed as
\begin{equation}
    \label{eq:D_sro}
    D_\t{sro} = D_\t{JS}(P \parallel P_\t{rss}),
\end{equation}
where $P_\t{rss}$ is the \textit{exact} motif population density distribution of a chemically random solid solution, as derived in ref.~\cite{sheriff2024chemicalmotif}. Larger values of $D_\t{sro}$ indicate a greater absolute amount of SRO in the state defined by $P$.

For annealing simulations, $P$ was evaluated using data from all seven snapshots across each of the 65 independent Monte Carlo runs described in section \ref{meth:3} (i.e., $7 \times 65 \times 4{,}394 = 1{,}999{,}270$ motifs per temperature). For TMP simulations, structures were first denoised using a score-based denoising algorithm\autocite{hsu_score-based_2024} for 8 steps. This was followed by identification of the crystal structure using the polyhedral template matching (PTM) algorithm with a root-mean-square deviation cutoff of 0.1. Only atoms identified as bcc were considered for constructing $P$. Minimally distorted motifs were mapped to ideal bcc motif geometry to ensure consistent motif classification. In both cases, $P$ was generated by randomly sampling subsets of 1,900,000 motifs to ensure consistency across datasets.

To determine whether SRO states obtained from TMP are distinguishable from equilibrium configurations, we adopted the methodology developed in ref.~\cite{islam_nonequilibrium_2025}. In this approach, the Jensen-Shannon divergence (eq.~\ref{eq:js}) between a nonequilibrium motif distribution $P$ (obtained from TMP simulations) and all equilibrium distributions $P_{\t{eq}}(T)$ (obtained from annealing simulations) are computed. The minimum value among these, denoted $D_\t{eff}$, represents the distance of a nonequilibrium state from its closest equilibrium counterpart. States with $D_\t{eff}$ that fall within their expected equilibrium baseline are classified as quasi-equilibrium states, whereas those that deviate are considered far-from-equilibrium states (see Supplementary Information section 4 for details).

Additional quantification of SRO was obtained using the widely employed Warren-Cowley (WC) parameters~\autocite{cowley_approximate_1950}, defined for each atomic pair $AB$ as:
\begin{equation}
    \label{eq:wc}
    \alpha_{AB} = 1 - \frac{p(A|B)}{c_{A}},
\end{equation}
where $p(A|B)$ is the conditional probability of finding an $A$ atom at a nearest-neighbor site of a $B$ atom, and $c_A$ is the average concentration of element $A$ in the alloy.


\section{Results}

\subsection{Thermomechanical processing-induced SRO}
\label{sec:1}

The stress-strain relationship for a representative processing condition ($T = 300\,\t{K}$ and $\dot{\epsilon} = 10^{9}\,\t{s}^{-1}$) of the TiTaVW TMP simulations is shown in fig.~\rref{figure_1}{b}. The stress-strain curve begins with a clear elastic region, where stress increases linearly with strain up to the yield stress, at which point the four initially seeded dislocation loops begin to move, initiating plastic deformation. Soon after yielding, the system reaches a plateau in which the flow stress fluctuates around a steady mean (i.e., steady-state flow stress). Aside from brief transients caused by changes in the loading direction, the steady flow stress remains consistent throughout the seven deformation cycles. A similar trend is observed in the evolution of dislocation density (fig.~\rref{figure_1}{c}), where the steady-state dislocation density $\rho^\infty$ is on the order of $10^{17}\,\t{m}^{-2}$. This value is consistent with previous molecular dynamics simulation results\autocite{zepeda-ruiz_probing_2017, zepeda-ruiz_atomistic_2021} and with theoretical estimates of the limiting dislocation density\autocite{cotterill_does_1977}. The steady behavior in both flow stress and dislocation density is observed across all temperature and strain rate conditions considered here (see Supplementary Information figs.~2 and 3), suggesting that it is a fundamental aspect of the alloy’s thermomechanical response, when dislocations serve as the primary carriers of plasticity\autocite{zepeda-ruiz_probing_2017}.

To examine how the SRO of the material evolves under such deformation, we monitored $D_{\t{sro}}$ (eq.~\ref{eq:D_sro}) as a function of strain, as illustrated in fig.~\rref{figure_1}{d}. With plastic deformation, the initially chemically random system forms SRO, evidenced by a rapid increase in $D_{\t{sro}}$ with strain. Remarkably, even at a strain of $0.3$, substantial SRO has already formed, indicating that modest levels of plastic strain --- typical in many materials processing conditions --- are sufficient to induce SRO in random solid solutions. This has important implications for interpreting the role of SRO on mechanical response during experiments: since SRO begins to evolve from the very onset of plastic deformation, even modest plastic strain can significantly alter the local chemical environment in the alloy\autocite{han_ubiquitous_2024}. As a result, any subsequent mechanical testing will reflect a modified SRO state, complicating efforts to isolate direct influence of SRO on mechanical properties.

With continued deformation, the system eventually reaches a steady value of $D_{\t{sro}}$, denoted $D^\infty_{\t{sro}}$, that is independent of the initial SRO state before deformation, as demonstrated in ref.~\cite{islam_nonequilibrium_2025}. This implies that even if we were to initialize the simulation with a more ordered configuration, such as a state annealed at moderate temperatures, the system would still evolve toward the same $D^\infty_{\t{sro}}$ shown in fig.~\rref{figure_1}{d}. The evolution of $D_{\t{sro}}$ with strain during TMP, across all processing conditions considered here (see Supplementary Information fig.~4), is well described by the following equation:
\begin{equation}
    \label{eq:sro}
    D_\t{sro}(\epsilon) = \left(D^{\t{o}}_\t{sro} - \frac{\Gamma}{\lambda}\right) \t{e}^{-\lambda \epsilon} + \frac{\Gamma}{\lambda}, 
\end{equation}
where $D^{\t{o}}_{\t{sro}}$ is the initial amount of SRO, and $\Gamma$ and $\lambda$ are rate parameters governing the SRO evolution kinetics. Note that $D^\infty_{\t{sro}} = D_\t{sro}(\epsilon \rightarrow \infty) = \Gamma/\lambda$, from eq.~\ref{eq:sro}. For the specific case shown in fig.~\rref{figure_1}{d} ($T = 300\,\t{K}$ and $\dot{\epsilon} = 10^{9}\,\t{s}^{-1}$), the values of $\Gamma$ and $\lambda$ are $2.02\times10^{-3}$ bits per unit strain and $1.03$ per unit strain, respectively.

Taking the derivative of eq.~\ref{eq:sro} with respect to strain yields the following SRO evolution equation:
\begin{equation}
\label{eq:sro_rate} \frac{\t{d}}{\t{d}\epsilon}(D_{\t{sro}}) = \Gamma - \lambda D_{\t{sro}}. 
\end{equation}
The two terms on the right-hand side of eq.~\ref{eq:sro_rate} reveal the existence of two competing processes governing the SRO evolution\autocite{islam_nonequilibrium_2025}. The first term ($\Gamma$) is positive and represents the rate of SRO creation, arising from the inherent chemical bias associated with dislocation motion. Driven by applied mechanical forces, dislocations preferentially traverse through low-energy pathways. The energetics of these pathways are dictated by the local chemical motifs near the dislocation core\autocite{utt_origin_2022, islam_nonequilibrium_2025}. As a result, dislocation motion in metallic alloys is chemically selective and promotes the formation of SRO by favoring certain motifs over others. The second term ($-\lambda D_{\t{sro}}$) is negative and proportional to the amount of SRO present in the system; it accounts for the destruction of SRO due to the randomizing effects of the mechanically-biased component of dislocation motion. Together, the SRO creation rate $\Gamma$ and annihilation rate $\lambda$ govern the overall evolution of SRO during TMP.

\begin{figure*}[!tb]
  \centering
  \includegraphics[width=\textwidth]{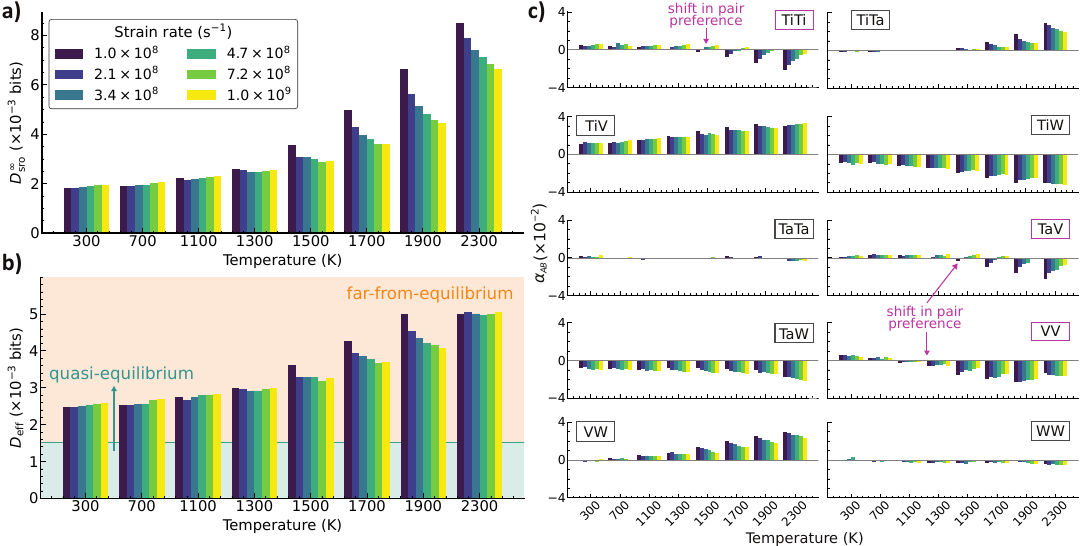}
  \caption{\label{figure_2} \textbf{Effect of temperature and strain rate on steady-state SRO.} \textbf{a)} Steady-state SRO ($D^\infty_{\t{sro}}$) as a function of temperature and strain rate. $D^\infty_{\t{sro}}$ values were obtained by averaging $D_{\t{sro}}$ from the seventh compression cycle. \textbf{b)} The steady-state SRO states are of far-from-equilibrium kind\autocite{islam_nonequilibrium_2025} and they move further from equilibrium as temperature increases, indicated by increasing $D_{\t{eff}}$ (sec.~\ref{meth:4}) \textbf{c)} Warren-Cowley parameters for different pairs indicate that there is a change in pair preference at moderate temperatures for certain elemental pairs (indicated by the purple box), marking a change in the chemical motifs forming the SRO.}
\end{figure*}

\subsection{Effects of processing conditions on steady-state SRO}
\label{sec:2}

We now turn our attention to the steady-states of SRO induced by TMP. Figure \rref{figure_2}{a} shows the steady-state SRO ($D^\infty_{\t{sro}}$) under a wide range of processing conditions. It is evident that both temperature and strain rate have systematic effects on $D^\infty_{\t{sro}}$. At low temperatures (300\,K--1,100\,K), $D^\infty_{\t{sro}}$ remains small, with only a weak positive correlation with temperature. Extrapolating this observation (sec.~\ref{sec:5}) to even lower temperatures approaching $0\,\t{K}$ shows that a finite degree of SRO persists, even in the absence of any thermal activation, due to chemical biases in the athermal and deformation-driven chemical mixing. Within this low-temperature regime, reducing the strain rate by an order of magnitude has little effect on $D^\infty_{\t{sro}}$. A stronger positive correlation between $D^\infty_{\t{sro}}$ and temperature emerges in the moderate temperature range ($1{,}300$–$1{,}500\,\t{K}$), where thermally activated dislocation mechanisms start to dominate. Beyond this regime, $D^\infty_{\t{sro}}$ increases rapidly with temperature. In this high-temperature regime, a pronounced strain rate dependence also emerges: slower deformation rates lead to larger $D^\infty_{\t{sro}}$. This trend indicates that chemical mixing driven by thermally-activated dislocation motion, when combined with slower mechanical driving (which allows more time for atomic rearrangement), promotes the formation of SRO.

Since TMP is a nonequilibrium process, the resulting SRO states are likewise nonequilibrium in nature. We quantify their deviation from equilibrium configurations using $D_{\t{eff}}$ (sec.~\ref{meth:4}), a metric that measures the statistical distance between a steady-state SRO configuration and its corresponding equilibrium counterpart using eq.~\ref{eq:js}. Larger $D_{\t{eff}}$ values indicate greater deviations from equilibrium. As shown in fig.~\rref{figure_2}{b}, $D_{\t{eff}}$ increases monotonically with temperature, implying that thermal activation of dislocation motion drives the system progressively further from equilibrium SRO. In equilibrium states, $D_{\t{eff}}$ values are expected to remain below $1.52\times10^{-3}$ bits (see Supplementary Information section 4). However, as indicated in fig.~\rref{figure_2}{b}, the steady-states of SRO obtained through TMP consistently exceed this threshold, confirming their far-from-equilibrium character. These far-from-equilibrium SRO states, observed during TMP, cannot be accessed via equilibrium pathways such as thermal annealing. As a result, TMP opens a new regime of SRO states beyond the traditional equilibrium space.

Finally, the modulation of steady-state SRO by TMP parameters is also reflected in their pairwise chemical preferences. Figure~\rref{figure_2}{c} shows the WC parameters (eq.~\ref{eq:wc}) for all atomic pairs as functions of temperature and strain rate. The overall trend parallels that of fig.~\rref{figure_2}{a}: increasing the temperature increases the degree of SRO. However, a particularly intriguing feature is observed --- namely, shifts in chemical pair preferences around 1,300\,K--$1{,}500\,\t{K}$. Specifically, Ti–Ti, Ta–V, and V–V exhibit distinct transitions across this temperature range, shifting from pair repulsion (positive WC parameters) to pair attraction (negative WC parameters). The temperature range over which this transition occurs aligns with the sharp rise in $D^\infty_{\t{sro}}$ with temperature seen in fig.~\rref{figure_2}{a}, indicating that the activation of thermal processes not only amplifies the amount of SRO but also alters its underlying chemical nature.

\subsection{Effects of processing conditions on SRO evolution rate}
\label{sec:3}

\begin{figure}[!tb]
  \centering
  \includegraphics[width=\columnwidth]{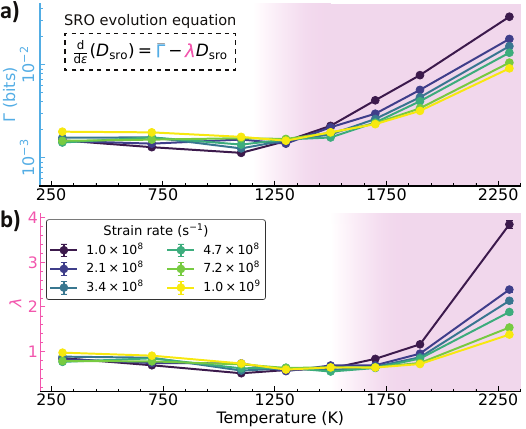}
  \caption{\label{figure_3} \textbf{Effect of temperature and strain rate on SRO evolution rate.} Temperature and strain rate dependence of \textbf{a)} SRO creation rate $\Gamma$, and \textbf{b)} SRO annihilation rate $\lambda$. The shaded purple region highlights the temperature range where both $\Gamma$ and $\lambda$ become sensitive to changes in temperature and strain rate. Error bars (one standard deviation) are obtained by propagating the covariance from the nonlinear fits of $\epsilon$–$D_{\mathrm{sro}}$ data to eq.~5. The error bars are} smaller than the marker size and therefore not visible.
\end{figure}

To investigate the effect of processing parameters on the evolution of SRO, we begin by examining how they influence the SRO creation ($\Gamma$) and annihilation ($\lambda$) rates introduced in section~\ref{sec:1}, which together govern SRO evolution rate with strain (eq.~\ref{eq:sro_rate}). The SRO creation rate $\Gamma$ as a function of strain rate and temperature is shown in fig.~\rref{figure_3}{a}. At the low-temperature regime, the effect of temperature and strain rate on $\Gamma$ is negligible. However, in the range of 1,100\,K–$1{,}500\,\t{K}$, we observe that $\Gamma$ increases rapidly with temperature. A strain rate dependence also emerges in this regime, with lower strain rates resulting in higher values of $\Gamma$. Notably, the temperature at which this transition --- from processing condition-insensitive to processing condition-sensitive behavior --- occurs appears to be itself strain-rate dependent, gradually shifting to lower temperatures as the strain rate decreases. The SRO annihilation rate ($\lambda$) exhibits a qualitatively similar trend, as shown in fig.~\rref{figure_3}{b}. However, the transition in $\lambda$ occurs at slightly higher temperatures, approximately in the range of 1,500\,K–$1{,}700\,\t{K}$.
\begin{figure*}[!bt]
  \centering
  \includegraphics[width=\textwidth]{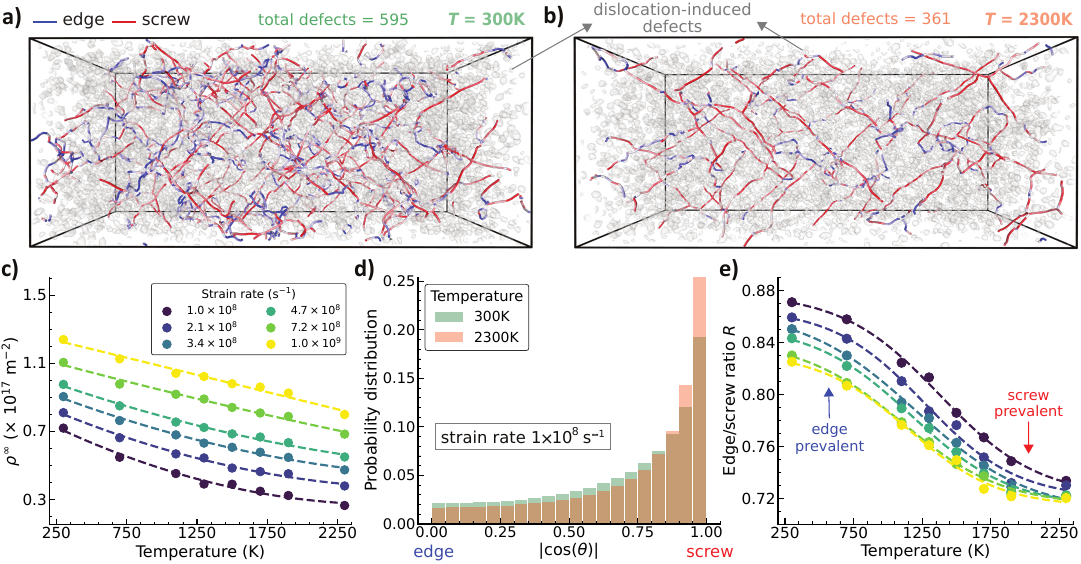}
  \caption{\label{figure_4} \textbf{Dislocation-mediated mechanisms of SRO evolution.} \textbf{a)} Dislocation configuration at $300\,\t{K}$ and \textbf{b)} $2300\,\t{K}$ after reaching steady state during thermomechanical processing at a strain rate of $10^8\,\t{s}^{-1}$. Grey features indicate dislocation-induced defects originating from cross-kinks (approximately 595 at 300\,K and 361 at 2,300\,K.) \textbf{c)} Steady-state dislocation density $\rho^\infty$ decreases with temperature and strain rate. $\rho^\infty$ values were obtained by averaging $\rho$ from the seventh compression cycle. \textbf{d)} Probability distribution of dislocation character at $\dot{\epsilon} = 10^{8}\,\t{s}^{-1}$, where $\theta$ is the angle between the Burgers vector and line direction $\theta$. \textbf{e)} Ratio of edge-to-screw character $R$ (eq.~\ref{eq:edge}) as a function of temperature and strain rate.}
\end{figure*}

While both $\Gamma$ and $\lambda$ increase with temperature beyond their respective transition points, their functional forms differ markedly. $\Gamma$ grows exponentially with temperature, increasing by one order of magnitude. In contrast, $\lambda$ increases approximately in a linear fashion, remaining within the same order of magnitude across the temperature range studied. This difference between $\Gamma$ and $\lambda$ quantitatively explains the rapid increase in steady-state SRO ($D^\infty_{\t{sro}} = \Gamma/\lambda$), observed at high temperatures (fig.~\rref{figure_2}{a}). Moreover, because $\lambda$ also appears in the exponential term of eq.~\ref{eq:sro}, its increase with temperature shortens the characteristic strain scale for SRO evolution, allowing the system to reach steady-state SRO more rapidly (i.e., over a smaller strain interval).

The dependence of the transition temperature of both $\Gamma$ and $\lambda$ on strain rate, observed in fig.~\ref{figure_3}, has important implications. Since control of SRO evolution via temperature becomes feasible only beyond the transition temperature, the ability to tune the transition temperature through strain rate provides a valuable degree of process control. Specifically, by sufficiently reducing the strain rate, the transition can be shifted to lower temperatures, thereby enabling SRO control under more accessible and experimentally achievable TMP conditions.

\subsection{Effects of processing conditions on SRO evolution mechanism}
\label{sec:4}

Having established that processing parameters can modulate the SRO creation and annihilation rates --- and consequently, SRO evolution --- we now turn to the underlying mechanisms responsible for this control. As explained in section~\ref{sec:1}, SRO formation during TMP occurs due to the chemical bias associated with dislocation motion\autocite{islam_nonequilibrium_2025}. Any phenomenon during deformation that enhances this chemical bias promotes SRO creation, while phenomena that counteract it contribute to SRO annihilation.

We first examine the role of steady-state dislocation density $\rho^\infty$ in modulating the chemical bias. A dense dislocation network restricts dislocation mobility through elastic interactions, geometrical constraints, and dislocation junctions. These constraints are evident in the dislocation configuration shown in fig.~\rref{figure_4}{a}, which corresponds to the processing condition of $T = 300\,\t{K}$ and $\dot{\epsilon} = 10^8\,\t{s}^{-1}$. The dislocations appear highly entangled, forming dense forests that inhibit their motion. In contrast, the dislocation configuration in fig.~\rref{figure_4}{b}, corresponding to $T = 2300\,\t{K}$ at the same strain rate, contains fewer dislocations, allowing for the formation of longer, more mobile dislocation lines. The grey features visible in both systems represent dislocation-induced defects such as vacancies, interstitials, and defect clusters. These defects originate primarily from the self-pinning of screw dislocations, a consequence of kink-pair nucleation on intersecting glide planes\autocite{zhou_cross-kinks_2021,marian_dynamic_2004}. Such defects are notably more prevalent in the system shown in fig.~\rref{figure_4}{a} than in fig.~\rref{figure_4}{b} (i.e., 64\% more defects at 300\,K than 2,300\,K), further impeding dislocation motion, especially of screw types. Together, these constraints limit the number of energetically distinct pathways accessible to dislocations, thereby suppressing the manifestation of chemical bias in their motion. As a result, \textit{elevated dislocation densities and defect concentrations reduce the extent of SRO formation}.

Figure~\rref{figure_4}{c} shows the variation of $\rho^\infty$ with strain rate and temperature. At low temperatures, $\rho^\infty$ is high, and, for a given temperature, higher strain rates result in higher $\rho^\infty$. Low temperatures also result in a larger number of dislocation-induced defects, as there is not enough thermal energy to overcome the cross-kinks. This explains the trends observed in fig.~\rref{figure_3}{a}, where the SRO creation rate $\Gamma$ increases with temperature. The origin of this increase lies, in part, in the reduction of $\rho^\infty$ (and associated defects) at elevated temperatures. Similarly, the enhancement of SRO formation at lower strain rates can be attributed to the corresponding decrease in $\rho^\infty$. Lower strain rates also provide more time for dislocations to sample available paths and preferentially select energetically favorable ones, further promoting chemically biased dislocation motion. The increased dislocation mobility at higher temperatures also contributes to the rise in the SRO annihilation rate $\lambda$ (fig.~\rref{figure_3}{b}) by enabling more frequent interactions between dislocations and chemical motifs. However, this increase is considerably less pronounced than the corresponding rise in $\Gamma$ (fig.~\rref{figure_3}{a}).

While the restrictive influence of dislocation forests and dislocation-induced defects accounts for much of the observed temperature and strain rate dependence, it does not fully explain the discernible transition in SRO evolution behavior at moderate temperatures (fig.~\rref{figure_3}) or the weak strain rate sensitivity observed at low temperatures. To resolve this, we examine the character of the dislocations present in the system and its temperature dependence. It is well established that at low temperatures, screw dislocations are considerably less mobile than their edge counterparts\autocite{yin_atomistic_2021} --- a difference that diminishes as temperature increases. The higher concentration of self-pinning features at lower temperatures (fig.~\rref{figure_4}{a}) is itself a manifestation of this behavior. In addition, fig.~\rref{figure_4}{d} shows that increasing temperature leads to a significant shift in the overall dislocation character toward screw. This phenomenon can be directly visualized in figs.~\rref{figure_4}{a} and \rref{figure_4}{b}, where it is clear that the system deformed at $300\,\t{K}$ possesses a larger fraction of edge dislocations compared to $2300\,\t{K}$. This overall shift toward a screw-dominated dislocation population at elevated temperatures, combined with their increased mobility, is mechanistically significant.

Among dislocation types, screw dislocations are more chemically biased in their motion. This is due to their ability to move via kink-pair formation and cross-slip. These mechanisms allow screw dislocations to sample numerous transition pathways, \textit{making their mobility more sensitive to the local chemical environment}. In contrast, edge dislocations are confined mostly to fixed glide planes, rendering their motion less dependent on chemistry (i.e., more mechanically driven). This distinction has been observed in ref.~\cite{yin_atomistic_2021}, where screw dislocation mobility was shown to depend sensitively on the local chemical ordering, whereas edge dislocation mobility remained largely unaffected.

Figure~\rref{figure_4}{e} shows a transition toward reduced ratio of edge-to-screw character $R$ (eq.~\ref{eq:edge}) at moderate temperatures. As $R$ decreases, the overall chemical bias of dislocation motion is amplified (due to decreased edge character), leading to enhanced SRO formation. This trend (along with the decrease in $\rho^\infty$) accounts for the observed increase in the SRO creation rate $\Gamma$ (fig.~\rref{figure_3}{a}) in this regime, causing it to outpace the annihilation rate $\lambda$ increase (fig.~\rref{figure_3}{b}). A particularly revealing aspect of the effect of dislocation character emerges when examining low-temperature conditions. Here, high strain rates produce two competing effects: they increase $\rho^\infty$ and associated defects (suppressing SRO creation), while simultaneously decreasing $R$ (promoting SRO creation). The net result of these opposing influences --- enhanced chemical bias from reduced edge character versus restricted motion from higher $\rho^\infty$ --- explains the weak strain rate dependence of SRO evolution at low temperatures seen in fig.~\ref{figure_3}. The high-temperature regime presents a different dynamic. As fig.~\rref{figure_4}{e} demonstrates, variations in strain rate have progressively less effect on $R$ at elevated temperatures. In this regime, $\rho^\infty$ remains the dominant factor: lower strain rates lead to reduced densities and, consequently, enhanced SRO formation.

The results discussed in this section reveal a strong coupling between dislocation character, dislocation density, and SRO evolution in equiatomic TiTaVW during TMP. $\rho^\infty$ values observed during TMP for the TiTaVW system are on the order of $10^{16}$–$10^{17}\ \t{m}^{-2}$, which is comparable to dislocation densities reported for other multi-principal element alloy systems such as HfNbTaTiZr\autocite{alhafez_nanoindentation_2024} ($3$–$5 \times 10^{16}\ \t{m}^{-2}$), CoCrFeNi\autocite{cao_dynamic_2023} ($0.9$–$2 \times 10^{17}\ \t{m}^{-2}$), FeCoCrNiCu\autocite{luo_microstructural_2021} ($\sim 6 \times 10^{16}\ \t{m}^{-2}$), and CrCoNi\autocite{cao_maximum_2022} ($6.7 \times 10^{16}$-$1.9 \times 10^{17} \,\t{m}^{-2}$) under similar deformation conditions. This consistency supports that the dislocation–SRO coupling mechanism identified here represents a general feature of compositionally complex alloys.

\subsection{Process-space extrapolation of steady-state SRO}
\label{sec:5}

\begin{figure*}[!tb]
  \centering
  \includegraphics[width=\textwidth]{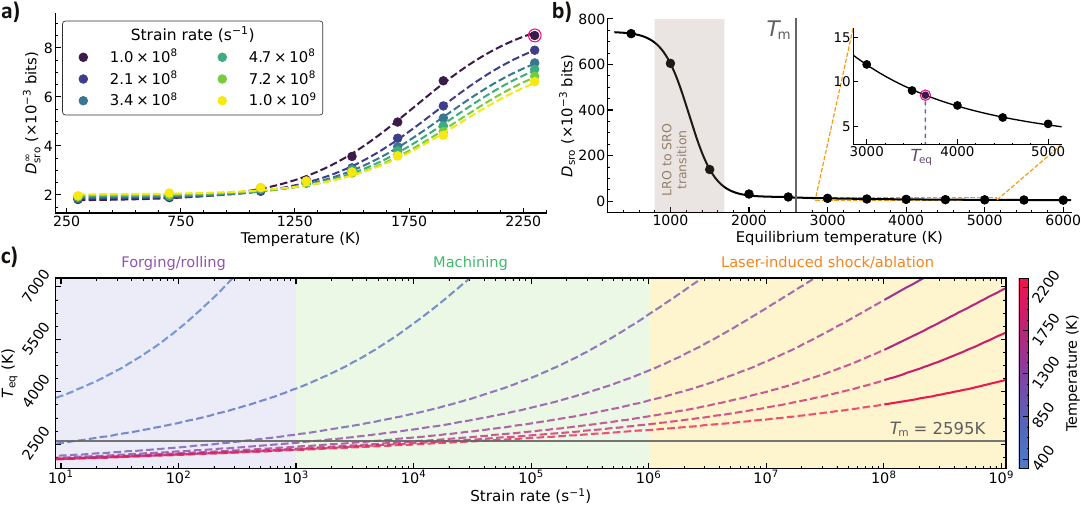}
  \caption{\label{figure_5} \textbf{Steady-state SRO during manufacturing processes.} \textbf{a)} The variation of steady-state SRO $D^\infty_{\t{sro}}$ induced by thermomechanical processing as a function of temperature and strain rate. The dashed lines show the empirical model (eq.~\ref{eq:empirical}) fit to the $T$-$D^\infty_{\t{sro}}$ data for each strain rate. \textbf{b)} Equilibrium SRO states obtained from annealing at different temperatures. The equivalent temperature, $T_{\t{eq}}$, is determined by matching $D^\infty_{\t{sro}}$ obtained from thermomechanical processing with its corresponding equilibrium SRO value. An example match is highlighted by the magenta circled markers; the corresponding $T_{\t{eq}}$ is indicated by the vertical dashed line in the inset.
 \textbf{c)} Map of $T_{\t{eq}}$ across a wide range of temperatures and strain rates, spanning three relevant manufacturing process categories. The dashed line segments indicate extrapolated values obtained from the empirical model fitted using $D^\infty_{\t{sro}}$ across all temperature and strain rate cases (see Supplementary Information section 5)}.
\end{figure*}

The processing conditions investigated here correspond to high strain-rate deformation, characteristic of processes such as laser-induced shock or ablation\autocite{righi_towards_2021,tang_quantifying_2024}. Nevertheless, our results can be extrapolated to predict steady-state SRO ($D^\infty_\t{sro}$) in lower strain-rate regimes more relevant to conventional materials processing. To enable such extrapolation we introduce the following empirical model:
\begin{equation}
\label{eq:empirical}
    D^\infty_\t{sro}(T, \dot{\epsilon}) = D^\t{ath}_\t{sro}(\dot{\epsilon}) + \frac{D^\t{sat}_\t{sro}(\dot{\epsilon}) - D^\t{ath}_\t{sro}(\dot{\epsilon})}{1 + \t{e}^{-w[T - T_\t{o}(\dot{\epsilon})]}}.
\end{equation}
Each parameter in this model has a well-defined physical interpretation that will be discussed next. As shown in fig.~\rref{figure_5}{a}, at the low-temperature limit, $D^\infty_\t{sro}$ reaches a finite minimum, indicating the persistence of steady-state SRO even in the absence of any thermal activation. We refer to this residual SRO as athermal SRO $D^\t{ath}_\t{sro}$. At the high-temperature end of fig.~\rref{figure_5}{a}, $D^\infty_\t{sro}$ approaches a saturation value that we call saturation SRO $D^\t{sat}_\t{sro}$. The transition temperature marking the onset of rapid SRO increase with temperature is denoted as $T_\t{o}$, while the parameter $w$ governs the sharpness of this transition. All model parameters, except for $w$, exhibit a linear dependence on the logarithm of strain rate (details in Supplementary Information section 5). Despite its simplicity, this empirical model provides an excellent fit to the simulation data (fig.~\rref{figure_5}{a}), enabling valid extrapolation to a broader range of processing conditions.

To place the steady-state SRO obtained from TMP in an equilibrium thermodynamic context, we map each state to its equilibrium counterpart based on SRO magnitude. This requires first understanding how the equilibrium SRO varies with temperature. As shown in fig.~\rref{figure_5}{b}, the equilibrium SRO exhibits a transition from long-range to short-range order between 1,000\,K-$1{,}500\,\t{K}$, followed by a gradual decrease with temperature (fig.~\rref{figure_5}{b} inset). Notably, the SRO never vanishes, even beyond the melting point ($T_{\t{m}} = 2595\,\t{K}$). For each thermomechanical condition, we define an equivalent temperature, $T_\t{eq}$, as the equilibrium temperature at which the equilibrium SRO most closely matches $D^\infty_{\t{sro}}$. This mapping, illustrated in fig.~\rref{figure_5}{b} inset, provides a thermodynamically meaningful framework for interpreting processing-mediated SRO.

Using the empirical model (eq.~\ref{eq:empirical}), we extrapolate $D^\infty_{\t{sro}}$ to slower strain rates and construct a corresponding map of $T_{\t{eq}}$ across a wide range of processing conditions in fig.~\rref{figure_5}{c}. The extrapolated strain-rate regimes are illustrated by three representative manufacturing process categories: forging/rolling\autocite{switzner_effect_2010,nwachukwu_effects_2017}, machining\autocite{jaspers_material_2002,zhang_strain_2020}, and laser-induced shock/ablation\autocite{righi_towards_2021,tang_quantifying_2024}. Figure~\rref{figure_5}{c} demonstrates that each of these categories can access a broad spectrum of SRO states, including states with $T_{\t{eq}}$ well below the melting point. For example, under a strain rate typical of forging/rolling ($20\,\t{s}^{-1}$) at a deformation temperature of $1100\,\t{K}$, the TiTaVW alloy can reach a steady-state SRO comparable in magnitude to the SRO state obtained by annealing at $2{,}200\,\t{K}$. However, as discussed in section~\ref{sec:2}, the chemical nature of these two SRO states would be fundamentally different.

It is important to note that our simulations were conducted at high strain rates, where diffusion-based mechanisms such as dislocation climb and point-defect migration are largely suppressed. Among these, vacancy migration is known to be responsible for SRO formation during equilibrium processes. As a result, the empirical model in eq.~\ref{eq:empirical} will not capture the full contribution of such diffusion-mediated mechanisms. At lower strain rates, longer timescales will allow these mechanisms to become active\autocite{xu_influence_2023}, potentially enhancing SRO further and consequently lowering $T_\t{eq}$ beyond the predictions of our model. Thus, the $T_\t{eq}$ values shown in fig.~\rref{figure_5}{c}, particularly at low strain rates, should be interpreted as conservative upper bounds estimates. In practice, even higher degrees of SRO may be attainable through TMP, which underscores the potential of TMP as a practical route for achieving targeted SRO states.

The presence of diffusion-based mechanisms at lower strain rates is expected to influence the parameters in eq.~\ref{eq:empirical}. Since the athermal SRO ($D^\t{ath}_\t{sro}$) represents the residual SRO at low temperature, this term is not expected to change appreciably once diffusion becomes active. In contrast, the saturation SRO ($D^\t{sat}_\t{sro}$) may increase as enhanced diffusion facilitates additional SRO formation on top of dislocation-mediated processes\autocite{xu_influence_2023}. The transition sharpness parameter, $w$, would likely become larger, reflecting the stronger Arrhenius-type sensitivity of diffusion-mediated SRO formation to temperature, whereas the onset temperature $T_\t{o}$ should remain largely unchanged.

Finally, the empirical model parameters introduced in eq.~\ref{eq:empirical}, as well as their dependence on $\dot{\epsilon}$ (see Supplementary Information fig.~6), are inherently material dependent. Both the thermodynamic driving force and the kinetics of SRO are governed by atomic size mismatch, electronegativity differences, and mixing enthalpy, all of which are expected to affect the model parameters. However, SRO evolution in the presence of both dislocation motion and diffusion is a complex phenomenon, and without a systematic exploration of these effects across different alloy systems and phases, it is difficult to anticipate their collective effect. Thus, a more comprehensive investigation of these dependencies would be a valuable direction for future work.


\section{Conclusion}

Building on our previous finding that dislocation motion both creates and destroys SRO, driving alloys into far-from-equilibrium states\autocite{islam_nonequilibrium_2025}, we demonstrate here that during TMP the evolution of SRO follows a quantitative balance between creation and annihilation rates, $\Gamma$ and $\lambda$, and we quantify how this balance varies with temperature and strain rate. We further explain this dependence through physical features of the dislocation network --- dislocation density, edge-to-screw ratio, and the population/annealing of cross-kink-induced defects --- which together govern the degree of chemically biased versus mechanically biased dislocation motion.

By systematically varying the processing conditions (temperature and strain rate), we identified two distinct regimes of SRO evolution. At low temperatures, high dislocation density and abundant dislocation-induced defects limit mobility and suppress SRO creation, yet the strain-rate dependence remains weak due to compensating changes in dislocation character. At moderate to high temperatures, thermally activated mechanisms reduce dislocation density and shift the dislocation population toward screw character, amplifying chemical bias and sharply increasing $\Gamma$ relative to $\lambda$. This mechanistic transition explains the strong temperature and strain rate sensitivity of steady-state SRO observed in this regime.

The steady-state SRO states reached under TMP are far-from-equilibrium\autocite{islam_nonequilibrium_2025}, lying outside the range accessible by thermal annealing and thus expanding the accessible SRO landscape. The framework established here --- linking processing parameters to $\Gamma$ and $\lambda$ through measurable dislocation properties --- provides a physically grounded, transferable basis for understanding and predicting SRO evolution in chemically complex alloys. A key limitation of the present study is that the results are derived from high strain rates atomistic simulations, which affects the direct applicability of the findings to lower, more practical deformation rates. However the mechanistic connections identified between dislocation density, character, and chemical bias are general, offering a route to integrate SRO evolution into broader models of microstructural development under TMP. Perhaps more importantly, the mechanistic basis for SRO evolution derived here is testable with diffraction, dislocation analysis, and chemical-correlation measurements during controlled TMP.

\subsection*{Data and code availability}

The software for SRO quantification can be found in our \texttt{ChemicalMotifIdentifier} Python package\autocite{cmi_github}. The machine learning potential can be found in our \texttt{Machine-learning-potentials-for-modeling-alloys- across-compositions} GitHub repository\autocite{gitrepo}. Our figure style is implemented in \texttt{LovelyPlots}\autocite{lovelyplots} under the \texttt{paper} style. 
Any custom code or data that is not currently available in these repositories can be subsequently added upon reasonable request to the corresponding author.

\subsection*{CRediT author statement}

\textbf{Mahmudul Islam:} Conceptualization, Methodology, Data Curation, Formal Analysis, Investigation, Software, Visualization, Writing – Original Draft.  
\textbf{Killian Sheriff:} Conceptualization, Methodology, Software, Formal Analysis, Investigation, Writing – Review \& Editing.  
\textbf{Rodrigo Freitas:} Conceptualization, Supervision, Project Administration, Funding Acquisition, Resources, Writing – Review \& Editing.

\subsection*{Acknowledgments}

This material is based upon work supported by the Air Force Office of Scientific Research (AFOSR) under award number FA9550-25-1-0199, through the Young Investigator Program. This work was also supported by the MathWorks Ignition Fund and MathWorks Engineering Fellowship Fund and the Exponent Fellowship. This research used resources of the Oak Ridge Leadership Computing Facility’s Frontier supercomputer at Oak Ridge National Laboratory, supported by the Office of Science of the U.S. Department of Energy under Contract No. DE-AC05-00OR22725, through an award from the INCITE program. Allocation of compute time was provided by the DOE Innovative and Novel Computational Impact on Theory and Experiment (INCITE) Program. The authors thank Vasily Bulatov and Nicolas Bertin of Lawrence Livermore National Laboratory for their valuable discussions.

\subsection*{Competing interests}

The authors declare no competing interests.

\clearpage
\printbibliography[heading=bibnumbered,title={References}]

@article{george_high-entropy_2019,
	title = {High-entropy alloys},
	doi = {10.1038/s41578-019-0121-4},
	journaltitle = {Nature Reviews Materials},
	author = {George, Easo P. and Raabe, Dierk and Ritchie, Robert O.},
	date = {2019-08},
}

@article{sheriff_quantifying_2024,
	title = {Quantifying chemical short-range order in metallic alloys},
	doi = {10.1073/pnas.2322962121},
	journaltitle = {Proceedings of the National Academy of Sciences},
	author = {Sheriff, Killian and Cao, Yifan and Smidt, Tess and Freitas, Rodrigo},
	date = {2024-06-18},
}

@article{sheriff2024chemicalmotif,
  title = {Chemical-motif characterization of short-range order with E(3)-equivariant graph neural networks},
  DOI = {10.1038/s41524-024-01393-5},
  journal = {npj Computational Materials},
  author = {Sheriff,  Killian and Cao,  Yifan and Freitas,  Rodrigo},
  year = {2024},
  month = sep,
}

@article{dasari_exceptional_2023,
	title = {Exceptional enhancement of mechanical properties in high-entropy alloys via thermodynamically guided local chemical ordering},
	doi = {10.1073/pnas.2211787120},
	journaltitle = {Proceedings of the National Academy of Sciences},
	author = {Dasari, Sriswaroop and Sharma, Abhishek and Jiang, Chao and Gwalani, Bharat and Lin, Wei-Chih and Lo, Kai-Chi and Gorsse, Stéphane and Yeh, An-Chou and Srinivasan, Srivilliputhur G. and Banerjee, Rajarshi},
	date = {2023-06-06},
}

@article{chen_simultaneously_2021,
	title = {Simultaneously enhancing the ultimate strength and ductility of high-entropy alloys via short-range ordering},
	doi = {10.1038/s41467-021-25264-5},
	journaltitle = {Nature Communications},
	author = {Chen, Shuai and Aitken, Zachary H. and Pattamatta, Subrahmanyam and Wu, Zhaoxuan and Yu, Zhi Gen and Srolovitz, David J. and Liaw, Peter K. and Zhang, Yong-Wei},
	date = {2021-08-16},
}

@article{zheng_multi-scale_2023,
	title = {Multi-scale investigation of short-range order and dislocation glide in {MoNbTi} and {TaNbTi} multi-principal element alloys},
	doi = {10.1038/s41524-023-01046-z},
	journaltitle = {npj Computational Materials},
	author = {Zheng, Hui and Fey, Lauren T. W. and Li, Xiang-Guo and Hu, Yong-Jie and Qi, Liang and Chen, Chi and Xu, Shuozhi and Beyerlein, Irene J. and Ong, Shyue Ping},
	date = {2023-05-30},
}

@article{qiu_corrosion_2017,
	title = {Corrosion of high entropy alloys},
	doi = {10.1038/s41529-017-0009-y},
	journaltitle = {npj Materials Degradation},
	author = {Qiu, Yao and Thomas, Sebastian and Gibson, Mark A. and Fraser, Hamish L. and Birbilis, Nick},
	date = {2017-08-21},
}

@article{xie_percolation_2021,
	title = {A percolation theory for designing corrosion-resistant alloys},
	doi = {10.1038/s41563-021-00920-9},
	journaltitle = {Nature Materials},
	author = {Xie, Yusi and Artymowicz, Dorota M. and Lopes, Pietro P. and Aiello, Ashlee and Wang, Duo and Hart, James L. and Anber, Elaf and Taheri, Mitra L. and Zhuang, Houlong and Newman, Roger C. and Sieradzki, Karl},
	date = {2021-06},
}

@article{xie_highly_2019,
	title = {Highly efficient decomposition of ammonia using high-entropy alloy catalysts},
	doi = {10.1038/s41467-019-11848-9},
	journaltitle = {Nature Communications},
	author = {Xie, Pengfei and Yao, Yonggang and Huang, Zhennan and Liu, Zhenyu and Zhang, Junlei and Li, Tangyuan and Wang, Guofeng and Shahbazian-Yassar, Reza and Hu, Liangbing and Wang, Chao},
	date = {2019-09-05},
}

@article{su_enhancing_2023,
	title = {Enhancing the radiation tolerance of high-entropy alloys via solute-promoted chemical heterogeneities},
	doi = {10.1016/j.actamat.2022.118662},
	journaltitle = {Acta Materialia},
	author = {Su, Zhengxiong and Ding, Jun and Song, Miao and Jiang, Li and Shi, Tan and Li, Zhiming and Wang, Sheng and Gao, Fei and Yun, Di and Ma, En and Lu, Chenyang},
	date = {2023-02-15},
}

@article{el_atwani_quinary_2023,
	title = {A quinary {WTaCrVHf} nanocrystalline refractory high-entropy alloy withholding extreme irradiation environments},
	doi = {10.1038/s41467-023-38000-y},
	journaltitle = {Nature Communications},
	author = {El Atwani, O. and Vo, H. T. and Tunes, M. A. and Lee, C. and Alvarado, A. and Krienke, N. and Poplawsky, J. D. and Kohnert, A. A. and Gigax, J. and Chen, W.-Y. and Li, M. and Wang, Y. Q. and Wróbel, J. S. and Nguyen-Manh, D. and Baldwin, J. K. S. and Tukac, O. U. and Aydogan, E. and Fensin, S. and Martinez, E.},
	date = {2023-05-02},
}

@article{moniri_three-dimensional_2023,
	title = {Three-dimensional atomic structure and local chemical order of medium- and high-entropy nanoalloys},
	doi = {10.1038/s41586-023-06785-z},
	journaltitle = {Nature},
	author = {Moniri, Saman and Yang, Yao and Ding, Jun and Yuan, Yakun and Zhou, Jihan and Yang, Long and Zhu, Fan and Liao, Yuxuan and Yao, Yonggang and Hu, Liangbing and Ercius, Peter and Miao, Jianwei},
	date = {2023-12},
}

@article{chen_direct_2021,
	title = {Direct observation of chemical short-range order in a medium-entropy alloy},
	doi = {10.1038/s41586-021-03428-z},
	journaltitle = {Nature},
	author = {Chen, Xuefei and Wang, Qi and Cheng, Zhiying and Zhu, Mingliu and Zhou, Hao and Jiang, Ping and Zhou, Lingling and Xue, Qiqi and Yuan, Fuping and Zhu, Jing and Wu, Xiaolei and Ma, En},
	date = {2021-04},
}

@article{zhang_short-range_2020,
	title = {Short-range order and its impact on the {CrCoNi} medium-entropy alloy},
	doi = {10.1038/s41586-020-2275-z},
	journaltitle = {Nature},
	author = {Zhang, Ruopeng and Zhao, Shiteng and Ding, Jun and Chong, Yan and Jia, Tao and Ophus, Colin and Asta, Mark and Ritchie, Robert O. and Minor, Andrew M.},
	date = {2020-05},
}

@article{schuh_mechanical_2015,
	title        = {Mechanical properties, microstructure and thermal stability of a nanocrystalline {CoCrFeMnNi} high-entropy alloy after severe plastic deformation},
	author       = {Schuh, B. and Mendez-Martin, F. and Völker, B. and George, E. P. and Clemens, H. and Pippan, R. and Hohenwarter, A.},
	doi          = {10.1016/j.actamat.2015.06.025},
	journaltitle = {Acta Materialia},
	date         = {2015-09-01},
}

@article{shahmir_effect_2016,
	title        = {Effect of annealing on mechanical properties of a nanocrystalline {CoCrFeNiMn} high-entropy alloy processed by high-pressure torsion},
	author       = {Shahmir, Hamed and He, Junyang and Lu, Zhaoping and Kawasaki, Megumi and Langdon, Terence G.},
	doi          = {10.1016/j.msea.2016.08.118},
	journaltitle = {Materials Science and Engineering: A},
	date         = {2016-10-31},
}

@article{cao_capturing_2024,
	title = {Capturing short-range order in high-entropy alloys with machine learning potentials},
	doi = {10.1038/s41524-025-01722-2},
	journaltitle = {npj Computational Materials},
	author = {Cao, Yifan and Sheriff, Killian and Freitas, Rodrigo},
	date = {2025-08-21},
}

@article{cao_maximum_2022,
	title        = {Maximum strength and dislocation patterning in multi–principal element alloys},
	author       = {Cao, Penghui},
	doi          = {10.1126/sciadv.abq7433},
	journaltitle = {Science Advances},
	date         = {2022-11-09},
}

@article{han_ubiquitous_2024,
	title = {Ubiquitous short-range order in multi-principal element alloys},
	doi = {10.1038/s41467-024-49606-1},
	journaltitle = {Nature Communications},
	author = {Han, Ying and Chen, Hangman and Sun, Yongwen and Liu, Jian and Wei, Shaolou and Xie, Bijun and Zhang, Zhiyu and Zhu, Yingxin and Li, Meng and Yang, Judith and Chen, Wen and Cao, Penghui and Yang, Yang},
	date = {2024-08-01},
}

@article{thompson_lammps_2022,
	title = {{LAMMPS} - a flexible simulation tool for particle-based materials modeling at the atomic, meso, and continuum scales},
	doi = {10.1016/j.cpc.2021.108171},
	journaltitle = {Computer Physics Communications},
	author = {Thompson, Aidan P. and Aktulga, H. Metin and Berger, Richard and Bolintineanu, Dan S. and Brown, W. Michael and Crozier, Paul S. and in 't Veld, Pieter J. and Kohlmeyer, Axel and Moore, Stan G. and Nguyen, Trung Dac and Shan, Ray and Stevens, Mark J. and Tranchida, Julien and Trott, Christian and Plimpton, Steven J.},
	date = {2022-02-01},
}

@article{zepeda-ruiz_probing_2017,
	title        = {Probing the limits of metal plasticity with molecular dynamics simulations},
	author       = {Zepeda-Ruiz, Luis A. and Stukowski, Alexander and Oppelstrup, Tomas and Bulatov, Vasily V.},
	doi          = {10.1038/nature23472},
	journaltitle = {Nature},
	date         = {2017-10},
}

@article{zepeda-ruiz_atomistic_2021,
	title        = {Atomistic insights into metal hardening},
	author       = {Zepeda-Ruiz, Luis A. and Stukowski, Alexander and Oppelstrup, Tomas and Bertin, Nicolas and Barton, Nathan R. and Freitas, Rodrigo and Bulatov, Vasily V.},
	doi          = {10.1038/s41563-020-00815-1},
	journaltitle = {Nature Materials},
	date         = {2021-03},
}

@article{stukowski_extracting_2010,
	title        = {Extracting dislocations and non-dislocation crystal defects from atomistic simulation data},
	author       = {Stukowski, Alexander and Albe, Karsten},
	doi          = {10.1088/0965-0393/18/8/085001},
	issn         = {0965-0393},
	journaltitle = {Modelling and Simulation in Materials Science and Engineering},
	date         = {2010-09},
}

@article{stukowski_visualization_2009,
	title        = {Visualization and analysis of atomistic simulation data with {OVITO}–the Open Visualization Tool},
	author       = {Stukowski, Alexander},
	doi          = {10.1088/0965-0393/18/1/015012},
	issn         = {0965-0393},
	journaltitle = {Modelling and Simulation in Materials Science and Engineering},
	date         = {2009-12},
}

@article{metropolis_equation_1953,
	title = {Equation of State Calculations by Fast Computing Machines},
	doi = {10.1063/1.1699114},
	journaltitle = {The Journal of Chemical Physics},
	author = {Metropolis, Nicholas and Rosenbluth, Arianna W. and Rosenbluth, Marshall N. and Teller, Augusta H. and Teller, Edward},
	date = {1953-06-01},
}

@article{hastings_monte_1970,
	title = {Monte Carlo Sampling Methods Using Markov Chains and Their Applications},
	doi = {10.2307/2334940},
	journaltitle = {Biometrika},
	author = {Hastings, W. K.},
	date = {1970},
}

@article{cowley_approximate_1950,
	title = {An Approximate Theory of Order in Alloys},
	doi = {10.1103/PhysRev.77.669},
	journaltitle = {Physical Review},
	author = {Cowley, J. M.},
	date = {1950-03-01},
}

@article{smidt_euclidean_2021,
	title        = {Euclidean Symmetry and Equivariance in Machine Learning},
	author       = {Smidt, Tess E.},
	doi          = {10.1016/j.trechm.2020.10.006},
	issn         = {2589-7209, 2589-5974},
	journaltitle = {Trends in Chemistry},
	date         = {2021-02-01},
}

@book{mackay2003information,
  title={Information theory, inference and learning algorithms},
  author={MacKay, David JC},
  year={2003},
  publisher={Cambridge university press},
}

@software{cmi_github,
	author       = {Sheriff, Killian and Cao, Yifan and Freitas, Rodrigo},
	url          = {https://github.com/killiansheriff/ChemicalMotifIdentifier}
}

@software{gitrepo,
	author       = {Sheriff, Killian and Freitas, Rodrigo},
	url          = {https://github.com/killiansheriff/Machine-learning-potentials-for-modeling-alloys-across-compositions}
}

@software{lovelyplots,
	title        = {LovelyPlots: A collection of matplotlib stylesheets for scientific figures},
	author       = {Sheriff, Killian},
	year         = 2023,
	month        = aug,
	publisher    = {Zenodo},
	doi          = {10.5281/zenodo.6903936}
}

@article{islam_nonequilibrium_2025,
	title = {Nonequilibrium chemical short-range order in metallic alloys},
	doi = {10.1038/s41467-025-64733-z},
	journaltitle = {Nature Communications},
	author = {Islam, Mahmudul and Sheriff, Killian and Cao, Yifan and Freitas, Rodrigo},
	date = {2025-10-08},
}

@article{zhou_atomic-scale_2022,
	title = {Atomic-scale evidence of chemical short-range order in {CrCoNi} medium-entropy alloy},
	doi = {10.1016/j.actamat.2021.117490},
	journaltitle = {Acta Materialia},
	author = {Zhou, Lingling and Wang, Qi and Wang, Jing and Chen, Xuefei and Jiang, Ping and Zhou, Hao and Yuan, Fuping and Wu, Xiaolei and Cheng, Zhiying and Ma, En},
	date = {2022-02-01},
}

@article{antillon_chemical_2021,
	title = {Chemical short range order strengthening in {BCC} complex concentrated alloys},
	doi = {10.1016/j.actamat.2021.117012},
	journaltitle = {Acta Materialia},
	author = {Antillon, E. and Woodward, C. and Rao, S. I. and Akdim, B.},
	date = {2021-08-15},
}

@article{antillon_chemical_2020,
	title = {Chemical short range order strengthening in a model {FCC} high entropy alloy},
	doi = {10.1016/j.actamat.2020.02.041},
	journaltitle = {Acta Materialia},
	author = {Antillon, E. and Woodward, C. and Rao, S. I. and Akdim, B. and Parthasarathy, T. A.},
	date = {2020-05-15},
}

@article{blades_tuning_2024,
	title = {Tuning chemical short-range order for stainless behavior at reduced chromium concentrations in multi-principal element alloys},
	doi = {10.1016/j.actamat.2024.120209},
	journaltitle = {Acta Materialia},
	author = {Blades, W. H. and Redemann, B. W. Y. and Smith, N. and Sur, D. and Barbieri, M. S. and Xie, Y. and Lech, S. and Anber, E. and Taheri, M. L. and Wolverton, C. and {McQueen}, T. M. and Scully, J. R. and Sieradzki, K.},
	date = {2024-09-15},
}

@article{yamanaka_stacking-fault_2017,
	title = {Stacking-fault strengthening of biomedical Co–Cr–Mo alloy via multipass thermomechanical processing},
	doi = {10.1038/s41598-017-10305-1},
	journaltitle = {Scientific Reports},
	author = {Yamanaka, Kenta and Mori, Manami and Sato, Shigeo and Chiba, Akihiko},
	date = {2017-09-07},
}

@article{sun_thermomechanical_2016,
	title = {Thermomechanical processing of metallic glasses: extending the range of the glassy state},
	doi = {10.1038/natrevmats.2016.39},
	journaltitle = {Nature Reviews Materials},
	author = {Sun, Yonghao and Concustell, Amadeu and Greer, A. Lindsay},
	date = {2016-06-07},
}

@article{rackwitz_understanding_2025,
	title = {Understanding thermo-mechanical processing pathways to simultaneously increase strength and damping in steels},
	doi = {10.1016/j.actamat.2025.120864},
	journaltitle = {Acta Materialia},
	author = {Rackwitz, J. and Olson, G. B. and Tasan, C. C.},
	date = {2025-05-01},
}

@article{trink_processing_2023,
	title = {Processing and microstructure–property relations of Al-Mg-Si-Fe crossover alloys},
	doi = {10.1016/j.actamat.2023.119160},
	pages = {119160},
	journaltitle = {Acta Materialia},
	author = {Trink, Bernhard and Weißensteiner, Irmgard and Uggowitzer, Peter J. and Strobel, Katharina and Hofer-Roblyek, Anna and Pogatscher, Stefan},
	date = {2023-09-15},
}

@article{fan_strain_2021,
	title = {Strain rate dependency of dislocation plasticity},
	doi = {10.1038/s41467-021-21939-1},
	journaltitle = {Nature Communications},
	author = {Fan, Haidong and Wang, Qingyuan and El-Awady, Jaafar A. and Raabe, Dierk and Zaiser, Michael},
	date = {2021-03-23},
}

@article{yin_atomistic_2021,
	title = {Atomistic simulations of dislocation mobility in refractory high-entropy alloys and the effect of chemical short-range order},
	doi = {10.1038/s41467-021-25134-0},
	journaltitle = {Nature Communications},
	author = {Yin, Sheng and Zuo, Yunxing and Abu-Odeh, Anas and Zheng, Hui and Li, Xiang-Guo and Ding, Jun and Ong, Shyue Ping and Asta, Mark and Ritchie, Robert O.},
	date = {2021-08-11},
}

@article{utt_origin_2022,
	title = {The origin of jerky dislocation motion in high-entropy alloys},
	doi = {10.1038/s41467-022-32134-1},
	journaltitle = {Nature Communications},
	author = {Utt, Daniel and Lee, Subin and Xing, Yaolong and Jeong, Hyejin and Stukowski, Alexander and Oh, Sang Ho and Dehm, Gerhard and Albe, Karsten},
	date = {2022-08-15},
}

@article{rizzardi_mild--wild_2022,
	title = {Mild-to-wild plastic transition is governed by athermal screw dislocation slip in bcc Nb},
	doi = {10.1038/s41467-022-28477-4},
	journaltitle = {Nature Communications},
	author = {Rizzardi, Q. and {McElfresh}, C. and Sparks, G. and Stauffer, D. D. and Marian, J. and Maaß, R.},
	date = {2022-02-23},
	
}

@article{kubilay_high_2021,
	title = {High energy barriers for edge dislocation motion in body-centered cubic high entropy alloys},
	doi = {10.1038/s41524-021-00577-7},
	journaltitle = {npj Computational Materials},
	author = {Kubilay, R. E. and Ghafarollahi, A. and Maresca, F. and Curtin, W. A.},
	date = {2021-07-19},
}

@article{lu_relative_2021,
	title = {Relative mobility of screw versus edge dislocations controls the ductile-to-brittle transition in metals},
	doi = {10.1073/pnas.2110596118},
	journaltitle = {Proceedings of the National Academy of Sciences},
	author = {Lu, Yan and Zhang, Yu-Heng and Ma, En and Han, Wei-Zhong},
	date = {2021-09-14},
}

@article{marian_dynamic_2004,
	title = {Dynamic transitions from smooth to rough to twinning in dislocation motion},
	doi = {10.1038/nmat1072},
	journaltitle = {Nature Materials},
	author = {Marian, Jaime and Cai, Wei and Bulatov, Vasily V.},
	date = {2004-03},
}

@article{zhou_hidden_2021,
	title = {The hidden structure dependence of the chemical life of dislocations},
	doi = {10.1126/sciadv.abf0563},
	journaltitle = {Science Advances},
	author = {Zhou, X. and Mianroodi, J. R. and Kwiatkowski da Silva, A. and Koenig, T. and Thompson, G. B. and Shanthraj, P. and Ponge, D. and Gault, B. and Svendsen, B. and Raabe, D.},
	date = {2021-04-16},
}

@article{hsu_score-based_2024,
	title = {Score-based denoising for atomic structure identification},
	doi = {10.1038/s41524-024-01337-z},
	journaltitle = {npj Computational Materials},
	author = {Hsu, Tim and Sadigh, Babak and Bertin, Nicolas and Park, Cheol Woo and Chapman, James and Bulatov, Vasily and Zhou, Fei},
	date = {2024-07-18},
}

@article{yin_ab_2020,
	title = {Ab initio modeling of the energy landscape for screw dislocations in body-centered cubic high-entropy alloys},
	doi = {10.1038/s41524-020-00377-5},
	journaltitle = {npj Computational Materials},
	author = {Yin, Sheng and Ding, Jun and Asta, Mark and Ritchie, Robert O.},
	date = {2020-07-29},
}

@article{VASP_1,
  title={Ab initio molecular dynamics for liquid metals},
  author={Kresse, Georg and Hafner, J{\"u}rgen},
  journal={Physical Review B},
  volume={47},
  number={1},
  pages={558},
  year={1993},
  publisher={APS},
  doi={10.1103/PhysRevB.47.558}
}

@article{VASP_2,
  title={Efficient iterative schemes for ab initio total-energy calculations using a plane-wave basis set},
  author={Kresse, Georg and Furthm{\"u}ller, J{\"u}rgen},
  journal={Physical Review B},
  volume={54},
  number={16},
  pages={11169},
  year={1996},
  publisher={APS},
  doi={10.1103/PhysRevB.54.11169}
}

@article{VASP_3,
  title={Efficiency of ab-initio total energy calculations for metals and semiconductors using a plane-wave basis set},
  author={Kresse, Georg and Furthm{\"u}ller, J{\"u}rgen},
  journal={Computational Materials Science},
  volume={6},
  number={1},
  pages={15--50},
  year={1996},
  publisher={Elsevier},
  doi={10.1016/0927-0256(96)00008-0}
}

@article{VASP_4,
  title={Ab initio molecular-dynamics simulation of the liquid-metal-amorphous-semiconductor transition in germanium},
  author={Kresse, Georg and Hafner, J{\"u}rgen},
  journal={Physical Review B},
  volume={49},
  number={20},
  pages={14251},
  year={1994},
  publisher={APS},
  doi={10.1103/PhysRevB.49.14251}
}

@article{VASP_PAW,
  title={From ultrasoft pseudopotentials to the projector augmented-wave method},
  author={Kresse, Georg and Joubert, Daniel},
  journal={Physical Review B},
  volume={59},
  number={3},
  pages={1758},
  year={1999},
  publisher={APS},
  doi={10.1103/PhysRevB.59.1758}
}

@article{PBE,
  title = {Generalized Gradient Approximation Made Simple},
  author = {Perdew, John P. and Burke, Kieron and Ernzerhof, Matthias},
  journal = {Phys. Rev. Lett.},
  volume = {77},
  issue = {18},
  pages = {3865--3868},
  numpages = {0},
  year = {1996},
  publisher = {American Physical Society},
  doi = {10.1103/PhysRevLett.77.3865}
}

@article{PAW,
  title = {Projector augmented-wave method},
  author = {Bl\"ochl, P. E.},
  journal = {Phys. Rev. B},
  volume = {50},
  issue = {24},
  pages = {17953--17979},
  numpages = {0},
  year = {1994},
  publisher = {American Physical Society},
  doi = {10.1103/PhysRevB.50.17953}
}

@article{lee_strength_2021,
	title = {Strength can be controlled by edge dislocations in refractory high-entropy alloys},
	doi = {10.1038/s41467-021-25807-w},
	journaltitle = {Nature Communications},
	author = {Lee, Chanho and Maresca, Francesco and Feng, Rui and Chou, Yi and Ungar, T. and Widom, Michael and An, Ke and Poplawsky, Jonathan D. and Chou, Yi-Chia and Liaw, Peter K. and Curtin, W. A.},
	date = {2021-09-16},
}

@article{freitas_machine-learning_2022,
	title = {Machine-learning potentials for crystal defects},
	doi = {10.1557/s43579-022-00221-5},
	journaltitle = {{MRS} Communications},
	author = {Freitas, Rodrigo and Cao, Yifan},
	date = {2022-10-01},
}

@article{hirth1983theory,
  title={Theory of dislocations},
  author={Hirth, John Price and Lothe, Jens and Mura, Toshio},
  journal={Journal of Applied Mechanics},
  year={1983},
  publisher={ASME International}
}

@article{vitek1968intrinsic,
  title={Intrinsic stacking faults in body-centred cubic crystals},
  author={Vitek, Vaclav},
  journal={Philosophical Magazine},
  year={1968},
  publisher={Taylor \& Francis}
}

@article{Zotov_2024,
doi = {10.1088/1361-651X/ad2d68},
date = {2024-03-01},
author = {Nikolay Zotov and Konstantin Gubaev and Julian Wörner and Blazej Grabowski},
title = {Moment tensor potential for static and dynamic investigations of screw dislocations in bcc Nb},
journal = {Modelling and Simulation in Materials Science and Engineering}
}

@misc{wang2022tamingscrewdislocationcores,
  title={The Taming of the Screw: Dislocation Cores in BCC Metals and Alloys}, 
  author={Rui Wang and Lingyu Zhu and Subrahmanyam Pattamatta and David J. Srolovitz and Zhaoxuan Wu},
  doi = {10.48550/arXiv.2209.12323},
  date={2022-09-21},
  publisher = {{arXiv}},
}

@article{cotterill_does_1977,
	title = {Does dislocation density have a natural limit?},
	doi = {10.1016/0375-9601(77)90321-8},
	journaltitle = {Physics Letters A},
	author = {Cotterill, R. M. J.},
	date = {1977-01-24},
}

@article{drautz_atomic_2019,
	title = {Atomic cluster expansion for accurate and transferable interatomic potentials},
	doi = {10.1103/PhysRevB.99.014104},
	journaltitle = {Physical Review B},
	author = {Drautz, Ralf},
	date = {2019-01-08},
}

@article{lysogorskiy_performant_2021,
	title = {Performant implementation of the atomic cluster expansion ({PACE}) and application to copper and silicon},
	doi = {10.1038/s41524-021-00559-9},
	journaltitle = {npj Computational Materials},
	author = {Lysogorskiy, Yury and Oord, Cas van der and Bochkarev, Anton and Menon, Sarath and Rinaldi, Matteo and Hammerschmidt, Thomas and Mrovec, Matous and Thompson, Aidan and Csányi, Gábor and Ortner, Christoph and Drautz, Ralf},
	date = {2021-06-28},
	note = {Publisher: Nature Publishing Group},
}

@article{jaspers_material_2002,
	title = {Material behaviour in metal cutting: strains, strain rates and temperatures in chip formation},
	doi = {10.1016/S0924-0136(01)01227-4},
	journaltitle = {Journal of Materials Processing Technology},
	author = {Jaspers, S. P. F. C and Dautzenberg, J. H},
	date = {2002-02-14},
}

@article{switzner_effect_2010,
	title = {Effect of forging strain rate and deformation temperature on the mechanical properties of warm-worked 304L stainless steel},
	doi = {10.1016/j.jmatprotec.2010.01.014},
	journaltitle = {Journal of Materials Processing Technology},
	author = {Switzner, N. T. and Van Tyne, C. J. and Mataya, M. C.},
	date = {2010-06-01},
}

@article{nwachukwu_effects_2017,
	title = {Effects of rolling process parameters on the mechanical properties of hot-rolled St60Mn steel},
	doi = {10.1016/j.cscm.2017.01.006},
	journaltitle = {Case Studies in Construction Materials},
	author = {Nwachukwu, Peter U. and Oluwole, Oluleke O.},
	date = {2017-06-01},
}

@article{zhang_strain_2020,
	title = {Strain Rate of Metal Deformation in the Machining Process from a Fluid Flow Perspective},
	doi = {10.3390/app10093057},
	journaltitle = {Applied Sciences},
	author = {Zhang, Keguo and Wang, Keyi and Liu, Zhanqiang and Xu, Xiaodong},
	date = {2020-01},
}

@article{righi_towards_2021,
	title = {Towards the ultimate strength of iron: spalling through laser shock},
	doi = {10.1016/j.actamat.2021.117072},
	journaltitle = {Acta Materialia},
	author = {Righi, Gaia and Ruestes, Carlos J. and Stan, Camelia V. and Ali, Suzanne J. and Rudd, Robert E. and Kawasaki, Megumi and Park, Hye-Sook and Meyers, Marc A.},
	date = {2021-08-15},
}

@article{tang_quantifying_2024,
	title = {Quantifying dislocation drag at high strain rates with laser-induced Microprojectile impact},
	doi = {10.1016/j.ijplas.2024.103924},
	journaltitle = {International Journal of Plasticity},
	author = {Tang, Qi and Hassani, Mostafa},
	date = {2024-04-01},

}

@article{xu_influence_2023,
	title = {Influence of short-range order on diffusion in multiprincipal element alloys from long-time atomistic simulations},
	doi = {10.1103/PhysRevMaterials.7.033605},
	journaltitle = {Physical Review Materials},
	author = {Xu, Biao and Ma, Shihua and Huang, Shasha and Zhang, Jun and Xiong, Yaoxu and Fu, Haijun and Xiang, Xuepeng and Zhao, Shijun},
	date = {2023-03-30},
}

@book{bulatov_computer_2006,
	title = {Computer Simulations of Dislocations},
	isbn = {978-0-19-852614-8},
	publisher = {{OUP} Oxford},
	author = {Bulatov, Vasily and Cai, Wei},
	date = {2006-11-02},
}

@article{zhou_cross-kinks_2021,
	title = {Cross-kinks control screw dislocation strength in equiatomic bcc refractory alloys},
	doi = {10.1016/j.actamat.2021.116875},
	journaltitle = {Acta Materialia},
	author = {Zhou, Xinran and He, Sicong and Marian, Jaime},
	date = {2021-06-01},
}

@misc{sheriff_machine_2025,
	title = {Machine learning potentials for modeling alloys across compositions},
	doi = {10.48550/arXiv.2506.12592},
	publisher = {{arXiv}},
	author = {Sheriff, Killian and Xiao, Daniel and Cao, Yifan and Owen, Lewis R. and Freitas, Rodrigo},
	date = {2025-06-14},
}

@inproceedings{atchley_frontier_2023,
	title = {Frontier: Exploring Exascale},
	doi = {10.1145/3581784.3607089},
	series = {{SC} '23},
	booktitle = {Proceedings of the International Conference for High Performance Computing, Networking, Storage and Analysis},
	publisher = {Association for Computing Machinery},
	author = {Atchley, Scott and Zimmer, Christopher and Lange, John and Bernholdt, David and Melesse Vergara, Veronica and Beck, Thomas and Brim, Michael and Budiardja, Reuben and Chandrasekaran, Sunita and Eisenbach, Markus and Evans, Thomas and Ezell, Matthew and Frontiere, Nicholas and Georgiadou, Antigoni and Glenski, Joe and Grete, Philipp and Hamilton, Steven and Holmen, John and Huebl, Axel and Jacobson, Daniel and Joubert, Wayne and Mcmahon, Kim and Merzari, Elia and Moore, Stan and Myers, Andrew and Nichols, Stephen and Oral, Sarp and Papatheodore, Thomas and Perez, Danny and Rogers, David M. and Schneider, Evan and Vay, Jean-Luc and Yeung, P. K.},
	date = {2023-11-11},
}

@article{luo_microstructural_2021,
	title = {Microstructural evolution and mechanical properties of {FeCoCrNiCu} high entropy alloys: a microstructure-based constitutive model and a molecular dynamics simulation study},
	doi = {10.1007/s10483-021-2756-9},
	journaltitle = {Applied Mathematics and Mechanics},
	author = {Luo, Gangjie and Li, Li and Fang, Qihong and Li, Jia and Tian, Yuanyuan and Liu, Yong and Liu, Bin and Peng, Jing and Liaw, P. K.},
	date = {2021-08-01},
}

@article{cao_dynamic_2023,
	title = {Dynamic deformation behaviors and mechanisms of {CoCrFeNi} high-entropy alloys},
	doi = {10.1016/j.actamat.2023.119343},
	journaltitle = {Acta Materialia},
	author = {Cao, Tangqing and Zhang, Qian and Wang, Liang and Wang, Lu and Xiao, Yao and Yao, Jiahao and Liu, Huaiyi and Ren, Yang and Liang, Jun and Xue, Yunfei and Li, Xiaoyan},
	date = {2023-11-01},
}

@article{alhafez_nanoindentation_2024,
	title = {Nanoindentation into a bcc high-entropy {HfNbTaTiZr} alloy—an atomistic study of the effect of short-range order},
	doi = {10.1038/s41598-024-59761-6},
	journaltitle = {Scientific Reports},
	author = {Alhafez, Iyad Alabd and Deluigi, Orlando R. and Tramontina, Diego and Merkert, Nina and Urbassek, Herbert M. and Bringa, Eduardo M.},
	date = {2024-04-20},
}

\end{document}